\documentclass[aps,twocolumn,prl,superscriptaddress,preprintnumbers,showpacs,tightenlines]{revtex4}
\usepackage{amssymb}
\usepackage{times}
\usepackage{graphicx}
\usepackage{epsfig}

\begin{document}
\title{General Quantum State Swap: an XY model analysis}
\author{Ben-Qiong Liu}
\affiliation{Key Laboratory of Cluster Science of Ministry of
Education, and School of Physics, Beijing Institute of Technology,
Beijing 100081, China}
\author{Lian-Ao Wu}
\affiliation{Department of Theoretical Physics and History of
Science, The Basque Country University (EHU/UPV), PO Box 644, 48080
Bilbao, Spain}
\affiliation{IKERBASQUE, Basque Foundation for
Science, 48011 Bilbao, Spain}
\author{Bin Shao$^1$}
\author{Jian Zou$^1$}
\pacs{03.67.Hk, 75.10.Pq}

\begin{abstract}
We consider an exact state swap, defined as the swap between two
quantum states $|A\rangle$ and $|B\rangle$ in the Hilbert space of a
quantum system. We show that, given an arbitrary Hamiltonian
dynamics, there is a straightforward approach to calculating the
probability of the occurrence of an exact state swap, by employing
an exchange operator $P_{AB}$. For a given dynamics, the
feasibilities of proposed quantum setups, such as quantum state
amplifications and transfers can be evaluated.  These setups are
only distinguished by different forms of $P_{AB}$, which easily lead
to innovative designs of quantum setups or devices. We illustrate
the method with the isotropic XY model, whose unnoticed features are
revealed.
\end{abstract}

\maketitle

\textit{Introduction.---} One of challenges in quantum control (QC) and quantum information processing (QIP) is
to reliably transmit known or unknown quantum states from one subspace to another in the whole Hilbert space of
a quantum system. The transmission may be from a state in one information processor $A$ to another in $B$ or between different energy levels, and can often be characterized by a swap between two remote or local states.  Examples include a swap between two neighbour or distant states, quantum-state amplification\cite{Lee,Cappellaro,Perez-Delgado,Kay}, quantum
entanglement transfer\cite{Banchi}, entanglement
routers\cite{Bayat}, or long distance quantum communications via
optical fiber. These swaps may be conveniently processed in terms of local spin couplings such as the XY interaction, where no dynamical control is required. One of these swaps, quantum state transfer, has been studied extensively. These studies have developed analytic and numerical methods to evaluate the feasibility of a given dynamics.

Typically, quantum state transfer through spin chains is achieved by
placing a spin state at one end of the chain and waiting for a
specific amount of time to let this state evolve naturally under
spin dynamics and propagate to the other end. When the quantum state
propagates across the spin chain, it often loses its integrity
\cite{Bose,Allen}.  The crucial task is to calculate the fidelity of
quantum transmission, which is defined by the overlap between the
received and expected states in the receiver,
$F=\sqrt{\langle\phi(0)|\rho(t)|\phi(0)\rangle}$. Here
$|\phi(0)\rangle$ is a state at the receiver with the same form as
the initial state, $\rho(t)$ is the reduced density matrix of the
receiver at time $t$ and is obtained by tracing over all but the
receiver's sites. However, this quantity cannot always be readily
obtained, and normally the complexity of numerical computations
grows with the distance between the sender and
receiver\cite{ZMWang}.

Here we introduce a general method to analyze the possibility of
exact quantum swaps for an arbitrary Hamiltonian $H$ and for
arbitrary time $\tau$. Without loss of generality, this paper will
focus on swaps between two separate processors $A$ and $B$ with the
same internal structure linked with media, exemplified by optical
lattices, Josephson junction arrays\cite{Wu2}. Specifically, we
construct an unitary operator $W$. The eigenstates of $W$ encode all
information of the possibilities of a specific state swap. The
problem of solving the Schr$\ddot{o}$dinger equation now becomes an
eigen-problem of $W$. More significantly, it can even become an
eigen-problem of these simple exchange operators under the
eigenstates of the Hamiltonian. The method is directly applicable to
swaps between two identical subspaces of a Hilbert space of an
arbitrarily-given quantum system.

It should be pointed out that while the perfect fidelity of state
transfer via a naturally-available interaction seems to be
unattainable as shown in \cite{Burgarth2005, Christandl,
BoseContemp, Nikolopoulos, Feder,Verstraete,Osborne,BurgarthBose},
it can be achieved by properly pre-engineering the coupling
strengths \cite{Christandl}. This process has been termed as perfect
state transfer and will exactly swap two states in processors $A$
and $B$.

\textit{Formalism.---}  Our first step is to employ the $A\Leftrightarrow B$ permutation operator
$E_{AB}$ to swap all states in processors $A$ and $B$, such
that the quantum information is transferred from $A$ to $B$, and vice verse. The
permutation operator can be expressed explicitly by
\begin{equation}
E_{AB}=\sum_{\alpha\beta}{(|\beta_A\rangle\langle\alpha_A|)\otimes(|\alpha_B\rangle\langle\beta_B|)},
\end{equation}
where $\alpha, \beta=1,2,\dots,2^K$ for $K$ qubits located in
processors $A$ and $B$ and $E_{AB}^2=1$.
$|\alpha(\beta)_{A(B)}\rangle$ refers to a state
$|\alpha(\beta)\rangle$ in processor $A(B)$. We then apply a gate
$V_B$ locally on processor $B$ to obtain a desired state
$|B\rangle$. The total operator for the two actions reads
\begin{equation}
P_{AB}=V_BE_{AB},
\end{equation}
with $P_{AB}P_{AB}^{\dagger}=1$. It can serve as a building block in innovative designs of quantum devices \cite{Wu1}. A quantum state transfer is the simplest case with $V_B=1$. The remote amplification
of a quantum state $|000\rangle_A$ to $|111\rangle_B$ can be
achieved by applying the permutation operator then followed by
$V_B=\sigma_x\sigma_x\sigma_x$. The remote-controlled exchange
$|01\rangle_A\to|10\rangle_B$ is realized by setting
$V_B=\exp{(-i\pi\sigma_x\sigma_x/4)}$.

We now introduce the above-mentioned operator
$W(\tau)=P_{AB}U(\tau)$, which is unitary
\begin{equation}
W^\dagger(\tau)W(\tau)=U^\dagger(\tau)P_{AB}^\dagger
P_{AB}U(\tau)=1.
\end{equation}
As any unitary operator, the operator $W(\tau)$ can be diagonalized
and has a complete set of orthonormal eigenvectors
$\{|\psi_m(0)\rangle\}_\tau$ and exponential eigenvalues
$\{\exp{(i\omega_m)}\}_\tau$, where $\omega_m$ are real. A vector $|\psi_m(0)\rangle$ in the
set obeys the eigenequation
\begin{equation}
W(\tau)|\psi_m(0)\rangle=\exp{(i\omega_m)}|\psi_m(0)\rangle.
\end{equation}
 It can be rewritten in a more interesting form
\begin{equation}
U(\tau)|\psi_m(0)\rangle=\exp{(i\omega_m)}P_{AB}|\psi_m(0)\rangle,
\end{equation}
where the unitary condition is used.  The left-hand side of Eq.(5)
is the wave function $|\psi_m(\tau)\rangle$ of the system initially
prepared at the eigenstate $|\psi_m(0)\rangle$. In the case that the
eigenstate $|\psi_m(0)\rangle$ is a product state
\begin{equation}
|\psi_m(0)\rangle=|A\rangle\otimes|C\rangle,
\end{equation}
where $|C\rangle$ denotes a state outside processor $A$, we can obtain
\begin{eqnarray}
|\psi_m(\tau)\rangle &=& \exp{(i\omega_m)}P_{AB}|A\rangle\otimes|C\rangle\nonumber\\
&=&\exp{(i\omega_m)}|B\rangle\otimes|C'\rangle,
\end{eqnarray}%
The use of those eigenstates $|\psi_m(0)\rangle$ as
the initial states will lead to the exact quantum
transmission $|A\rangle\Leftrightarrow|B\rangle$. The details
of states $|C\rangle$ and $|C'\rangle$ play no roles in this process.
This is an ideal situation and usually only happens for special families of
Hamiltonians as in the perfect state transfers.

Of particular interest is that, for a given Hamiltonian initially
prepared in any state $|\phi(0)\rangle$, our method can obtain the
probability of achieving exact quantum swaps
$|A\rangle\Leftrightarrow|B\rangle$. The prescription is the
numerical diagonalization of the operator $W(\tau)$ such that we can
calculate the overlap between the initial state $|\phi(0)\rangle$
and the eigenstates $|\psi_m(0)\rangle$, which is the probability of
exact quantum swap, $p_m=|\langle\phi(0)|\psi_m(0)\rangle|^2$. If
the overlap $|\langle\phi(0)|\psi_m(0)\rangle|^2\to1$,  an exact
quantum transmission or swap occurs.  The swap is partial when $1>
|\langle\phi(0)|\psi_m(0)\rangle|^2>1/2 $. This overlap is a direct
indicator of quantum quantum swaps between $A$ and $B$.

\bigskip \textit{A case analysis: The isotropic XY model.---} Spin
chains are of great interest in quantum information science since
they are natural candidates for quantum channels in atomic scales.
The sender can transfer a quantum state to
the receiver via a naturally available Hamiltonian and does not require
manipulation or control over the chains. The Hamiltonian of
the isotropic Heisenberg XY chain reads
\begin{equation}
H=-J\sum_{j=1}^N(S_j^xS_{j+1}^x+S_j^yS_{j+1}^y),
\end{equation}
where the uniform interaction strength $J$ between nearest neighbour
sites is taken as $J=1$ for simplicity, and $S_j^\gamma
(\gamma=x,y,z)$ are the spin-half operators at the $j$th lattice
site. $N$ is the total number of spins and is assumed to be odd for
convenience. The periodic boundary conditions
$(S_{N+1}^\gamma=S_1^\gamma)$ are used. In addition, we consider the
whole system in the "one-magnon" state, in which the number of
spin-ups in the chain is one.
Besides, the conclusions for the XY model in the subspaces of the
zero and one magnons are applicable for the XXZ spin chain
$H=J\sum_{j=1}^N{(S_j^xS_{j+1}^x+S_j^yS_{j+1}^y+\Delta
S_j^zS_{j+1}^z+hS_j^z)}$ in the subspace of zero and one magnons .

The Hamiltonian (8) can be diagonalized via a Jordan-Wigner map
followed by a Fourier transformation. The eigenvalues and
eigenstates are \cite{Antal}
\begin{equation}
E_m=-\cos(\frac{2\pi m}{N}),
|\Phi_m\rangle=\frac{1}{\sqrt{N}}\sum_j{e^{i2\pi
mj/N}|\widehat{j}\rangle},
\end{equation}
where $m=1,\cdots,N$, the state $|\widehat{j}\rangle=|00\cdots
j\cdots 0\rangle$ represents that the state of the $j$th site has
been flipped to a spin-up state while the other spins remain
spin-down and spans the one-magnon subspace.

Here we will study several cases, where we calculate the probability
of the exact quantum transmission or swap. In the first case, the
exchange operator $P_{AB}$ swaps the states of corresponding spin
pairs in the chain, corresponding to quantum state transfer.

\textit{A. Quantum state transfer.---} In view of the great
potentialities of solid-state quantum information, focus has been on
implementation of quantum state transfer by spin chains. We now
start with the initial state $|\phi(0)\rangle=|\widehat{1}\rangle$
and let this state propagate to the $r$th spin after time $\tau$,
where $r=(N+1)/2$ is the site of our receiver. The exchange operator
$P_1=|\widehat{1}\rangle\langle
\widehat{r}|+|\widehat{r}\rangle\langle\widehat{1}|+P_0$ as depicted
in Fig.1(a), where
$P_0=\sum_j{|\widehat{j}\rangle\langle\widehat{j}|}$ ($j\ne 1,r$).
Using the eigenstates (9), we can numerically obtain $N$ eigenstates
and eigenvalues of $W(\tau)=P_1U(\tau)$, which are functions of
$\tau$, to seek possible values of  $\tau$ such that
$p_m=|\langle\psi_m(0)|\widehat{1}\rangle|^2\to 1$. Ideally, if we
find that $|\langle\psi_m(0)|\widehat{1}\rangle|^2=1$ and
$|\langle\psi_m(0)|\widehat{j}\rangle|^2=0$ (for $j\ne 1$) at a time
$\tau$, we have a perfect transfer for the state $|1\rangle$ from
the first site to the $r$th site at time $\tau$. However, our
numerical calculations run over all eigenstates and long time period
and show that there is no exact state transmission for the XY model,
as expected.

\begin{table}[htbp]
\begin{tabular}{llllllll}
\hline
$\tau$ &0.07 &0.13 &0.17 &0.19 &0.38 &0.44 &0.52\\
\hline
$\psi_m$ &$\psi_1$ &$\psi_1$ &$\psi_1$ &$\psi_2$ &$\psi_3$ &$\psi_5$ &$\psi_7$\\
\hline
$p_m$ &0.4997 &0.4989 &0.4982 &0.4977 &0.4910 &0.4879 &0.4837\\
\hline\\
\end{tabular}
\caption{The maximal $p_m$ and corresponding values of $\tau$. The
corresponding eigenvalues are $e^{i\omega_m}\approx-1$.}
\end{table}

On the other hand, the conventional quantum state transfer refers to
the process of transferring an unknown state, which requires at
least two eigenstates of $W$. Ideally, if
$\{|\psi_0(0)\rangle\}_\tau$ and $\{|\psi_1(0)\rangle\}_\tau$ can
both perform exact state transmissions, an unknown state
$a\{|\psi_0(0)\rangle\}_\tau+b\{|\psi_1(0)\rangle\}_\tau$ can be
transferred perfectly if
$\Delta\omega=\omega_0(\tau)-\omega_1(\tau)=2\pi K$, with $K$ being
arbitrary integers. In the XY model, for instance, the unknown state
can be $|\phi(0)\rangle=a|\widehat{0}\rangle+b|\widehat{1}\rangle$,
where $|\widehat{0}\rangle$ is the zero-magnon state and a trivial
eigenstate of $W(\tau)$. This is equivalent to the state
$a|0\rangle+b|1\rangle$ encoded at the first site initially. We now
define a joint probability of unknown state transfers,
$F=\langle\phi(0)|(a|\widehat{0}\rangle+be^{i\omega_m}|\psi_m\rangle)$,
explicitly $F=a^2+b^2e^{i\omega_m}\langle\widehat{1}|\psi_m\rangle$.
This joint probability may be surprisingly high, even if the
probability $p_m$ is about $0.5$, for instance, $F\approx0.97$ when
$a=\sqrt{\frac{9}{10}}$, $b=\sqrt{\frac{1}{10}}$, and $F\approx0.85$
even if $a=b=\frac{\sqrt{2}}{2}$. When $a>b$, the dominate
contribution in this joint probability $F$ is the zero-magnon state,
which can always propagate from the sender to the receiver with the
probability 1. Table \uppercase\expandafter{\romannumeral1} lists
the maximum probability $p_m$ ($m=1,2,\dots,N$) and corresponding
time interval $\tau$ for the exchange operator $P_1$.

\begin{figure}
\centering
\includegraphics[width=6.0cm]{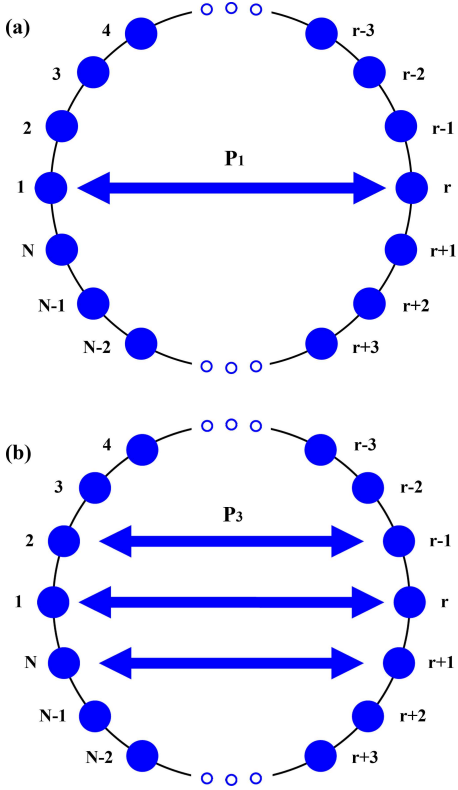}
\hspace{0.5cm}
\caption{Schematic of our quantum transmission
protocol: (a)$P_1$ and (b) $P_3$. The circles represent qubits.}
\end{figure}

We can also consider multi-qubit sender and receiver, for example 3
qubits as depicted in Fig.1(b), where the exchange operator is
$P_3=|\widehat{1}\rangle\langle
\widehat{r}|+|\widehat{2}\rangle\langle\widehat{r-1}|+|\widehat{N}\rangle\langle\widehat{r+1}|+H.c.+P_0'$,
with $P_0'=\sum_j{|\widehat{j}\rangle\langle\widehat{j}|}$, ($j\ne
1,2,r-1,r,r+1,N$). This operator swaps three pairs of spins while
keeping other sites intact. An exact quantum swap means
$p_3=\sum_{j=1,2,N}{|\langle\psi_m(0)|\widehat{j}\rangle|^2}=1$,
where the state of processor $A$ composed of three spins (1st, 2nd,
and $N$th) would ideally propagate to the targeted processor $B$.
Our numerical calculation shows that the results of $p_3$ remain
almost the same as those in one-qubit processor.

It is a conjecture that the probability of an exact quantum
transmission might be improved when the number of spin pairs in
processors is increased. We construct the exchange operator that
swaps all the spin sites in pairs
$P_{All}=\big(\sum_1^{M_1}{|\widehat{j}\rangle\langle\widehat{r+1-j}|}+\sum_{r+1}^{M_2}{|\widehat{j}\rangle\langle\widehat{N+r+1-j}|}\big)+H.c.$,
where $M_1=(N+1)/4$ and $M_2=(3N-1)/4$ when $(N-3)/2$ is even, while
$M_1=(N-1)/4$, $M_2=(3N+1)/4$ when $(N-3)/2$ is odd. We calculate
the probability $\sum_1^r{|\langle\psi_m(0)|\widehat{j}\rangle|^2}$
as a function of chain size $N$ in Fig.2. The probability decays
with the numbers of spins. Our numerical calculations show that the
probabilities seem to increase few percent for a given site number
in one-qubit, three-qubit or multi-qubit processors but not much.

\begin{figure}
\centering
\includegraphics[width=8.0cm]{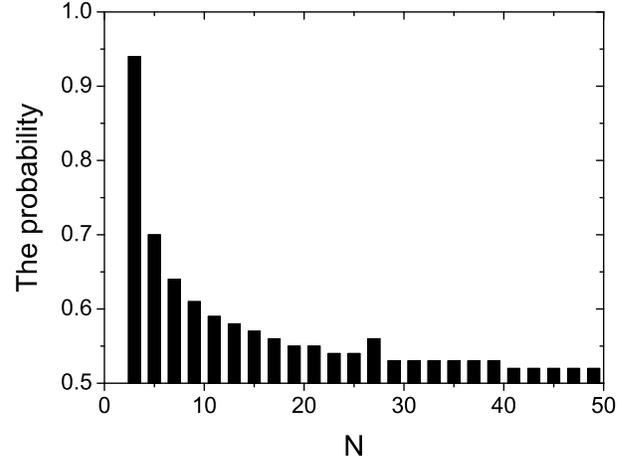}
\hspace{0.5cm}
\caption{The probability of an exact quantum
transmission for $P_{All}$ as a function of chain size $N$.}
\end{figure}

Although the isotropic spin-half XY model seems not to be good
candidate for accessing exact quantum transmissions or swaps, it is
interesting to note that it may be good enough for perfect state
transfer when $a>b$ holds. In other words, the XY model seems to a
good candidate of high-quality quantum state transfer for a family
of \textit{unknown} states, in particular states with $a>b$.

\textit{B. Remote entanglement.---} The exchange operator can also help to study remote entanglement
through the spin chain, where the receiver's spins are entangled while the sender's were not.

Consider the exchange operator
$P_E=\frac{1}{\sqrt{2}}|\widehat{1}\rangle(\langle\widehat{r}|+\langle\widehat{r+1}|)+H.c.+P_0''$,
which makes the two spins $r$th and $(r+1)$th in the maximally
entangled state $\frac{\sqrt{2}}{2}(|10\rangle+|01\rangle)$, with
$P_0''=\sum_j{|\widehat{j}\rangle\langle\widehat{j}|}$,
($j\ne1,r,r+1$). Our numerical calculation shows that the maximum
probability of exact remote entanglement is always $\sim0.5$, and
almost does not decay with the increase of $N$. We present the
probability $|\langle\phi(0)|\psi_1(0)\rangle_\tau|^2$ in Fig.3(a).
It seems that the probability can be still high for a long chain
(e.g., when $N=73$, $p\approx0.4995$), in fact, there always exist
at least one eigenstate $|\psi_m(0)\rangle$ ($m\in[1,N]$) which can
have an overlap $\sim0.5$ with $|\phi(0)\rangle$, independent of the
chain length. This result is interesting and shows that the
entangling ability of the XY model seems not to decay with the
distance, which should imply an unnoticed quantumness in this model.
In addition, this character of the XY model revealed here may
effectively support the point of view that high fidelity state
transfer over a long chain is possible\cite{Allen}.

Remote entanglement can also take place between remote pairs of
qubits. In Fig.3(b), we take the exchange operator
$P'_E=\frac{1}{\sqrt{2}}|\widehat{2}\rangle(\langle\widehat{1}|+\langle\widehat{r}|)+H.c.+P^\star_0$
which has the two spins 1st and $r$th maximally entangled, with
$P^\star_0=\sum_j{|\widehat{j}\rangle\langle\widehat{j}|}$
($j\ne1,2,r$), and calculate the overlap
$|\langle\phi(0)|\psi_3(0)\rangle_\tau|^2$. For the case $N=3$, the
overlap is relatively high, $p_{max}=0.73$, so that it will give
surprisingly high joint probability, for instance, $F\approx0.96$
when $a=\frac{\sqrt{3}}{2}$. Note that except for the eigenstate
marked by the subscript $m=3$, there are another $N-1$ choices for
us in order to find the maximal overlap with the initial state
$|\phi(0)\rangle$. We have numerically computed the overlaps between
each eigenstate of $W(\tau)$ and the initial state, and find again
that the probability 0.5 can always be achieved, no matter how large
the chain is.

\begin{figure}
\centering
\includegraphics[width=8.6cm]{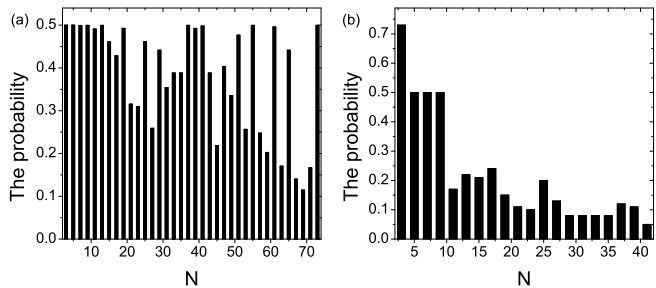}
\hspace{0.5cm}
\caption{The probability of an exact quantum
transmission for (a) $P_E$, (b) $P'_E$ as a function of chain size
$N$.}
\end{figure}

Similarly, we can generalize the use of the exchange operator to
realize the \textit{entanglement spreading} by the definition
$P_{ES}=\frac{1}{\sqrt{N}}|\widehat{1}\rangle(\langle\widehat{2}|+\cdots+\langle
\widehat{N}|)+H.c.$

\textit{Conclusions.---}We have introduced a general numerical
method for exact quantum transmissions, demonstrating that given an
arbitrary Hamiltonian at an arbitrary time $\tau$, the unitary
operator $W(\tau)=P_{AB}U(\tau)$ can be easily numerically
diagonalized such that we can obtain the probability of an exact
quantum swap by calculating the overlap between the initial state of
the whole system $|\phi(0)\rangle$ and the eigenstates of $W(\tau)$.
We find that high-quality quantum state transfers may be possible
for a family of unknown states using the XY model. The probability
for this model to create entanglement remotely does not decay with
the size of the spin chain, an interesting quantum feature. In
principle, it should be a fundamental feature of closed-system
quantum dynamics. Experience gained from the XY model sheds light on
the numerical method and will allow one to assess future directions
for other applications.

This work is supported by the National Natural Science Foundation of
China under Grant Nos.11075013 and 10974016, the Ikerbasque
Foundation Start-up, the Basque Government (Grant IT472-10), and the
Spanish Ministerio de Educaci\'on y Ciencia (MEC)(Project
No.FIS2009-12773-C02-02).


\end{document}